\newcommand{\be}{\begin{equation}}
\newcommand{\ee}{\end{equation}}
\newcommand{\bea}{\begin{eqnarray}}
\newcommand{\eea}{\end{eqnarray}}
\newcommand{\sect}{\section}

\newcommand{\ep}{\epsilon}
\newcommand{\ga}{\gamma}

\newcommand{\de}{\delta}

\newcommand{\non}{\nonumber}

\newcommand{\ba}{\begin{array}}
\newcommand{\ea}{\end{array}}
\documentstyle[12pt,twoside]{article}
\input amssym.def
\include{psfig}
\begin{document}
\vspace{-1mm}
\begin{flushright} G\"{o}teborg ITP 97-04\phantom{i} \\
\end{flushright}
\vspace{1mm}
\begin{center}{\bf\Large\sf BRST and Anti BRST structure of a topological
current algebra}
\end{center}
\vspace{1mm}
\begin{center}{{\bf\large Bani Mitra S\"{o}dermark{\normalsize
\footnote{bani@fy.chalmers.se}}\vspace{5mm}}\\{\em Institute of Theoretical
Physics, Chalmers Institute of Technology, S-412 96 G\"{o}teborg, Sweden}}
\end{center}
\begin{abstract}
The BRST structure of a current satisfying a non abelian affine algebra in two
dimensions was\
 studied by Isidro and Ramallo \cite{Isidro1}
 and an $N=2$ Superconformal Algebra was obtained.\
In this paper, we study the total BRST and anti BRST structure of the
topological algebra. We\
end up with an $N=4$ Superconformal algebra in which the central charge drops
out of most of\
the OPE's. The price one has to pay is that the no. of operators proliferates
tremendously\
and the algebra becomes infinite dimensional.
\end{abstract}
\setcounter{equation}{0}
\sect{Introduction}
The study of topological quantum field theories  \cite{Witten1,Birmingham} has
aroused great interest since their inception. These theories possess a
nilpotent symmetry $Q$ satisfying $ Q^2=0 $ and the physical observables are
the cohomology classes of the operator $Q$ as it acts in the full Hilbert
space. Such theories are therefore also known as cohomological field theories.
Since the stress energy tensor $ T_{\alpha \beta} $ is $Q$ exact,i.e.
\be
     T_{\alpha \beta}= [Q,G_{\alpha \beta}]
\ee
correlation functions of the observables are independent of the two dimensional
metric $ g_{\alpha \beta}$. The topological theories therefore have a much
larger symmetry than conformal theories. However, even in the category of
topological quantum field theory, there is a counterpart of conformal
invariance, since the special class of models in which $ T_{\alpha \beta} $ is
traceless, even before restricting to the $Q$ cohomology. These are the
topological conformal field theories (TCFT's).
In the same way that the conformal theories correspond to critical points in
the space of ordinary quantum field theories (QFT's) the topological CFT's are
the critical points in the space of topological QFT's. One can consider
perturbations of a TCFT by turning on the couplings of the appropriate physical
fields. This maintains the topological symmetry, but in general destroys the
conformal invariance, so one obtains a parameter family of more general
'massive' topological theories.
      The study of two dimensional TCFT's \cite{Witten1,Birmingham} has
attracted much interest due to its connection with non-critical string theories
\cite{Witten2,Dijkgraaf,Spiegelglas,Hu,Sadov}. A possible way of generating new
TCFT's is to study the BRST structure of different chiral algebras that extend
the Virasaro algebra. This method has been applied in \cite{Isidro1} and
\cite{Figueroa} to the case of an affine Lie Algebra, whereas in
\cite{Isidro2}, the analysis was extended to the case of a superconformal
current algebra. In \cite{Ennes}, the analysis was extended to the case of an
arbitrary affine Lie superalgebra.
      The topological symmetry of a TCFT is encoded in its topological algebra
which is the operator algebra closed by the chiral algebra of the TCFT and the
BRST current. It was shown in \cite{Isidro1,Isidro2,Ennes}  that the
topological algebra of a TCFT possessing a nonabelian current algebra symmetry
must include operators of dimensions one, two and three, realizing the so
called Kazama algebra \cite{Kazama}. This algebra differs from the standard
twisted $ N = 2 $ SCA as the former includes two dimension three operators and
can be regarded as an extension of the latter. The operator algebra of
\cite{Figueroa} is, however similar to that of \cite{Eguchi}.
     The anti BRST symmetry, introduced in 1976 by Curci and Ferrari
\cite{Curci} and Ojima \cite{Ojima} has been studied extensively by Baulieu et
al \cite{Baulieu}, Hwang \cite{Hwang} and Perry et al \cite{Perry} among
others. The question that has most often been asked is whether the anti BRST
condition $ \bar
{Q} |phys> = 0 $ yields any further information, in addition to the imposition
of the BRST condition $ Q|phys> = 0 $. The answer to the above questions have
been studied in detail as well \cite{Perry,Preitschopf,Kulshreshtha,Fulop}.
    It is the purpose of this paper to investigate the consequence of anti BRST
invariance in TCFT's, in particular, to examine the SCA formed by the
generators of the BRST and the anti BRST symmetry, thus extending the algebra
found in \cite{Isidro1}. The most important feature of our results is that the
central charge drops out almost totally, except in the Operator Product
Expansion (OPE) of the conformal charges with each other. The price we have to
pay for this is that the no. of terms in the algebra increases, for each
dimension, and the algebra can, in fact, be shown to be infinite dimensional.
       \\
    The paper is organized as follows: In Sec.2, we investigate the effects of
the BRST and anti BRST transformations on the matter and ghost fields and see
that they satisfy the conditions of a topological algebra. We next obtain the
algebra of the generators and see that the central charge drops out of most of
the equations. In Sec. 3, we see how the algebra becomes infinite dimensional,
while in Sec. 4, we present an analysis of the operators of the algebra of
order one, upto dimension 3. In Sec. 5, we conclude by summing up the results
of our analysis and some additional comments.

\setcounter{equation}{0}
\sect{The Topological Algebra}\
{Let us consider a Lie Algebra ${\bf g}$ generated by the Hermitian matrices $
T^{a}\, (a = 1,.. \dim{\bf g} )$ that satisfy the commutation relations
\be
     [T^{a},T^{b}] = if^{abc} T^{c}
\ee
The Lie Algebra ${\bf g}$ is assumed to be semisimple so that
\be
      f^{abc} f^{dbc} = c_{A} \de^{ad}
\ee
$c_{A}$  being the quadratic Casimir of the adjoint representation of the
Algebra. A holomorphic current taking values on $ {\bf g } $ is an operator $
J_{a}(z) $
with Laurent modes defined by
\be
      J_{a}(z) = \sum_{n \in {\Bbb Z} } J_{a}^{n}z^{-n-1}
\ee
The OPE of two currents is given by
\be
    J_{a}(z_{1})J_{b}(z_{2}) =\frac{1}{(z_{1}-z_{2})^2} k \de_{ab} + \frac{1}
{z_{1}-z_{2}} if^{abc}J_{c}
\ee
  The Sugawara energy momentum tensor for the currents $ J_{a} $ is given by
\cite{Fuchs}
\be
         T(z) = \frac{1}{(2k+c_{A})} J_{a}(z) J_{a}(z)
\ee
It is important to mention here that all expressions are assumed to be normal
ordered unless specified to the contrary and that only singular terms appear in
the OPE's of the operators.
It can be checked using standard methods that
\bea
        T(z_{1}) J_{a}(z_{2}) &=& \frac{J_{a}(z_{2})}{(z_{1}-z_{2})^2} +
\frac{\partial J_{a}}{(z_{1}-z_{2})} \\
        T(z_{1}) T(z_{2}) &=& \frac{c}{2(z_{1}-z_{2})^4} +
\frac{2T(z_{2})}{(z_{1}-z_{2})^2} + \frac{\partial T(z_{2})}{(z_{1}-z_{2})}
\eea

where the central charge in (2.7) is given by
\be
        c = \frac{2kdim{\bf g}}{2k + c_{A}}
\ee
In order to define a topological theory, we need to define a BRST symmetry. In
order to do this, we introduce the Faddeev-Popov ghosts, $  \ga_{a}(z_{1}) $
and $ \rho_{a}(z_{1}) $ and anti ghosts $ \bar {\ga_{a}}(z_{1}) $ and $ \bar
{\rho_{a}}(z_{2}) $ :
\bea
        \ga_{a}(z_{1}) \rho_{b}(z_{2}) &=& -\frac{\de_{ a b }}{(z_{1}-z_{2})}
\\   \bar {\ga_{a}}(z_{1}) \bar {\rho_{b}}(z_{2}) &=& -\frac{\de_{ a b
}}{(z_{1}-z_{2})}
\eea
We use the standard expression for the BRST charge $Q$ (see for example)
\cite{Hwang} for a constrained Lagrangian $ {\cal L} $. In this case, the
currents are the constraints, as for example in the gauged WZNW model
\cite{Kar} . We introduce the Lagrangian multipliers, $ v_{a} $ and their
conjugate multipliers $ P_{b} $:
\be
      v_{a}(z_{1}) P_{b}(z_{2}) = \frac{\de_{ a b }}{z_{1}-z_{2}}
\ee
The expression for the BRST charge takes the following form
\be
   Q = -:\ga_{a}J_{a}: - \frac{i}{2} f_{abc}:\ga_{a}\ga_{b}\rho_{c}: -
:P_{a}\bar {\rho_{a}}:
\ee
  The expression used above differs from that of \cite{Isidro1} and
\cite{Figueroa} due to the inclusion of the last term. Our intention was to see
if the algebra resulting from the inclusion of the Lagrangian multipliers would
be nontrivially different. The answer to the above question is a resounding yes
as we shall see in the following sections.
It is easy to check that the BRST charge introduces the following
transformation on $ J_{a} $ and $ \ga_{a} $
\bea
   {\de} J_{a} &=& if_{abc}:\ga_{b}J_{c}: - k{\partial \ga_{a}}  \\ {\de}
\ga_{a} &=& if_{abc}:\ga_{b}\ga_{c}:
\eea
and that \be  {\de}^2 \ga_{a} = {\de}^2J_{a} = 0 \ee
due to the Jacobi Identity.
The BRST transformation introduced on $ \rho_{a} $ is given by
\be
 {\de}\rho_{a} = J_{a} + if_{abc}: \ga_{b}\rho_{c}:
\ee
Due to the nilpotency of the BRST charge, $ {\de}^2{\rho} $ must vanish. We see
that
\be
{\de}^2{\rho} = -k{\partial \ga_{a}}
-\frac{f_{abc}f_{bmn}}{2}::\ga_{m}\ga_{n}:{\rho_{c}}: +
f_{abc}f_{cmn}:\ga_{b}\ga_{m}{\rho_{n}}:
=
\ee
The last term on the right takes the form
\bea
  f_{abc}f_{cmn}:\ga_{m}{\rho_{n}}{\ga_{b}}: - f_{abc}f_{cmn}
\partial\ga_{m}{\de_{bn}} & = & -f_{abc}f_{cmn}(\ga_{m}\ga_{b}{\rho_{n}}) -
{\de}_{bn} \partial \ga_{m}f_{abc}f_{cmb}
\eea
on using the identity \cite{Isidro1}
\be
   [\ga_{b}^{n}, :(\ga_{r}{\rho_{s}})^m:] = {\de}_{sb}\ga_{r}^{n+m}
\ee
so that the right hand side of (2.17) takes the following form:
\be
  -(k + c_{A}) \partial\ga_{a}
\ee
while the other terms drop out because of the Jacobi Identity.
Hence the BRST transformation is nilpotent on $ {\rho_{a}} $ only for the
critical level $ k = -c_{A} $.
The BRST transformation on the antighost fields yield
\bea
{\de}\bar {\rho_{a}} &=& 0;\\ {\de}\bar \ga_{a} &=& P_{a}; \\ {\de}P_{a} &=&
0;\\ {\de}v_{a} &=& \bar {\rho}_{a}
\eea
and ${\de}^2$ is easily seen to vanish on all of them.
We now consider the action of the anti BRST transformation. The anti BRST
operator takes the form \cite{Hwang}
\be
\bar {Q} = -\bar \ga_{a}J_{a} - \frac{i}{2}f_{abc}\bar \ga_{a}\bar \ga_{b}\bar
{\rho}_{c} + P_{a}{\rho}_{a} - if_{abc}\bar \ga_{a}\ga_{b}{\rho}_{c} -
if_{abc}\bar\ga_{a}v_{b}P_{c}
\ee
We notice the presence of the following currents in the expressions for $Q$ and
$\bar {Q}$
:
\bea
J^{a}_{gh} &=& if_{abc}\ga_{b}{\rho}_{c}; \\ J^{a}_{agh} &=& if_{abc}\bar
\ga_{b}\bar {\rho_{c}}; \\ J^{a}_{Lag} &=& if_{abc}v_{b}P_{c};
\eea
with levels given in Table 1.
\begin{table}
\caption{Levels of the Anti BRST Currents}
\begin{center}
\begin{tabular}{|l|r|}     \hline
Current & Level \\    \hline
$J_{a}$ & $-c_{A}$  \\   \hline
$J_{a}^{gh}$ & $c_{A}$  \\ \hline
$J_{a}^{agh}$ & $c_{A}$  \\ \hline
$J_{a}^{Lag}$ &  $-c_{A}$  \\ \hline
\end{tabular}
\end{center}
\end{table}
The anti BRST transformations on the fields $J_{a}$, $\ga_{a}$,${\rho}_{a}$,
$\bar\ga_{a}$, $\bar{\rho}_{a}$, $v_{a}$ and $P_{a}$ are as follows:
\bea
\bar{\de}J_{a} &=& if_{abc}\bar\ga_{b}J_{c} - k{\partial \bar\ga_{a}} ;\\
\bar{\de}\ga_{a} &=& if_{abc}\bar\ga_{b}\ga_{c} -P_{a} ;\\
\bar{\de}{\rho_{a}} &=& if_{abc}\bar\ga_{b}{\rho_{c}} ;\\
\bar{\de}\bar\ga_{a} &=& i\frac{f_{abc}}{2}\bar\ga_{b}\bar\ga_{c}    ;\\
\bar{\de}\bar{{\rho}_{a}} &=& J_{a} + if_{abc}\bar\ga_{b}\bar\rho_{c} +
if_{abc}\ga_{b}{\rho_{c}} + if_{abc}v_{b}P_{c} = {\cal T}_{a}   ;\\
\bar{\de}v_{a} &=& -{\rho_{a}} + if_{abc}\bar\ga_{b}v_{c}  ;\\
\bar{\de}P_{a} &=& if_{abc}\bar\ga_{b}P_{c};
\eea
It is easy to check using the Jacobi Identity that $\bar{\de}$ is nilpotent on
all of the fields except for $\bar{\rho}_{a}$, where it is nilpotent for
\be
k = -c_{A}
\ee
Let us look at the algebra formed by ${\cal J}_{a}$ and ${\cal T}_{a}$. After a
short calculation, we see that
\bea
{\cal J}_{a}(z_{1}){\cal J}_{b}(z_{2}) &=&
\frac{(k+c_{A})}{(z_{1}-z_{2})^2}{\de}_{ab} + \frac{if_{abc}{\cal
J}_{c}(z_{2})}{(z_{1}-z_{2})}  \\
{\cal T}_{a}(z_{1}){\cal T}_{b}(z_{2}) &=&
\frac{(k+c_{A})}{(z_{1}-z_{2})^2}{\de}_{ab} + \frac{if_{abc}{\cal
T}_{c}(z_{2})}{(z_{1}-z_{2})}
\eea
and for $k = -c_{A}$, the algebras formed by $({\cal J}_{a}, {\rho_{a}})$ and
$({\cal T}_{a},\bar{\rho_{a}})$ takes the form of the topological current
algebra introduced in \cite{Isidro1}.
\bea
{\cal J}_{a}(z_{1}){\cal J}_{b}(z_{2}) &=& if_{abc}\frac{{\cal
J}_{c}(z_{2})}{z_{1}-z_{2}} \\
{\cal J}_{a}(z_{1}){\rho_{b}}(z_{2})&=&
if_{abc}\frac{{\rho}_{c}(z_{2})}{(z_{1}-z_{2})}  \\
{\rho_{a}}(z_{1}){\rho_{b}}(z_{2}) &=& 0
\eea
and
\bea
{\cal T}_{a}(z_{1}){\cal T}_{b}(z_{2}) &=& if_{abc}\frac{{\cal
T}_{c}(z_{1})}{(z_{1}-z_{2})} \\
{\cal T}_{a}(z_{1})\bar{\rho}_{b}(z_{2}) &=&
if_{abc}\frac{\bar{\rho}_{c}(z_{2})}{(z_{1}-z_{2})} \\
\bar{\rho}_{a}(z_{1})\bar{\rho}_{b}(z_{2}) &=& 0
\eea
The OPE between the other elements of the algebra $({\cal
J}_{a},{\rho}_{a},{\cal T}_{a},\bar{\rho}_{a})$ are as follows:
\bea
 {\cal T}_{a}(z_{1}){\rho}_{b}(z_{2}) &=&
if_{abc}\frac{{\rho}_{c}(z_{2})}{(z_{1}-z_{2})}\\
{\cal J}_{a}(z_{1}){\cal T}_{b}(z_{2}) &=& if_{abc}\frac{{\cal
J}_{c}(z_{2})}{(z_{1}-z_{2})} \\
{\cal J}_{a}(z_{1})\bar{\rho}_{b}(z_{2}) &=& 0 \\
{\rho}_{a}(z_{1})\bar{\rho}_{b}(z_{2}) &=& 0
\eea\
The absence of a central term in the algebra formed by $({\cal J}_{a},
{\rho}_{a}, {\cal T}_{a}, \bar{\rho}_{a})$ is noted, as is the fact that
$({\cal J}_{a},{\rho}_{a})$ constitutes an ideal of the total algebra formed by
$ ({\cal J}_{a}, {\rho}_{a}, {\cal T}_{a}, \bar{\rho}_{a}) $ .
   We now attempt to construct the total energy momentum tensor.
Following \cite{Isidro1}, we consider the BRST variation of
\be
{\de}[\frac{{\rho}_{a}J_{a} + v_{a}\partial\ga_{a}}{2k+c_{A}}] =
\frac{J_{a}J_{a}}{2k+c_{A}} + \frac{k{\rho}_{a}\partial\ga_{a}}{2k+c_{A}} +
\bar{\rho}_{a}\partial\bar\ga_{a} + v_{a}\partial P_{a}
\ee
The right hand side is the energy-momentum tensor of the $({\rho},\ga) $ system
if the coefficient of the ${\rho}\partial\ga $ term is one, which happens for $
k = -c_{A} $. We note that for this value of $k$, the total energy momentum
tensor $T$ of the system is obtained. Hence the BRST partner $G$ of $T$ is
given by
\be
G = -\frac{{\rho}_{a}J_{a}}{c_{A}} + v_{a}\partial\bar\ga_{a}
\ee
We consider the anti BRST variation of
\be
\bar G = \frac{\bar{\rho}_{a}J_{a}+if_{abc}v_{b}\ga_{c}J_{a}}{2k+c_{A}} -
v_{a}\partial \ga_{a}
\ee
We obtain
\bea
\bar\de [\frac{\bar{\rho}_{a}J_{a}}{2k+c_{A}} +
i\frac{f_{abc}v_{b}\ga_{c}J_{a}}{2k+c_{A}} -v_{a}\partial\ga_{a}] \
&=& \frac{J_{a}J_{a}}{2k+c_{A}} + k\frac{\bar{{\rho}}_{a}\partial
\bar\ga_{a}}{2k+c_{A}} + {\rho}_{a}\partial\ga_{a} + v_{a}\partial P_{a} \non
\\ &&
 + ikf_{abc}\frac{v_{b}\ga_{c}\partial \bar\ga_{a}}{2k+c_{A}} -\
 if_{abc}v_{b}\ga_{c}\partial \bar\ga_{a}
\eea
This is the total energy momentum tensor for $ k = -c_{A} $.
We see thus at this critical value of $k$, that $T$ is both BRST and anti BRST
exact, and thus the theory is topological. We find it convenient to specify the
combination of terms $\bar{\rho}_{a} + if_{abc}v_{b}\ga_{c}$ as $R_{a}$. Hence,
\be
\bar G_{a} = -\frac{R_{a}J_{a}}{c_{A}} - v_{a}\partial\ga_{a}
\ee
We now attempt to look at the algebra formed by the generators $Q$, $G$, $\bar
Q$,  $\bar G$ , and $T$. It may be checked after a little calculation that
\bea
Q(z_{1})Q(z_{2}) &=& 0   \\
\bar Q(z_{1}) \bar Q(z_{2}) &=& 0   \\
Q(z_{1})\bar Q(z_{2}) &=& 0
\eea
We also have,
\bea
T(z_{1})Q(z_{2}) &=& \frac{Q(z_{2})}{(z_{1}-z_{2})^2} + \frac{\partial
Q(z_{2})}{(z_{1}-z_{2})}    \\
T(z_{1})\bar Q(z_{2}) &=& \frac{\bar Q(z_{2})}{(z_{1}-z_{2})^2} +
\frac{\partial \bar Q(z_{2})}{(z_{1}-z_{2})}  \\
T(z_{1})G(z_{2}) &=& \frac{2G(z_{1})}{(z_{1}-z_{2})^2}+ \frac{\partial
G(z_{2})}{(z_{1}-z_{2})}    \\
T(z_{1})\bar G(z_{2}) &=& 2\frac{\bar G(z_{2})}{(z_{1}-z_{2})^2} +
\frac{\partial \bar G(z_{2})}{(z_{1}-z_{2})}
\eea
and
\be
T(z_{1})T(z_{2}) = 2\frac{T(z_{2})}{(z_{1}-z_{2})^2} + \frac{\partial
T(z_{2})}{(z_{1}-z_{2})}
\ee
{\em so the total central term of the system vanishes}. This is a nontrivial
result and we feel it is a consequence of the fact that the full algebra with
all the Lagrangian multipliers is being considered.
We also have
\bea
Q(z_{1})G(z_{2}) &=& \frac{R_{gh}(z_{2})}{(z_{1}-z_{2})^2} +
\frac{T(z_{1})}{(z_{1}-z_{2})} \\
\bar{Q}(z_{1}) \bar{G}(z_{2}) &=& \frac{R_{agh}(z_{2})}{(z_{1}-z_{2})^2} +
\frac{T(z_{2})}{(z_{1}-z_{2})}
\eea
where $R_{gh}$ and $R_{agh}$ are defined as
\bea
R_{gh} &=& {\rho}_{a}\ga_{a} + P_{a}v_{a}   \\
R_{agh} &=& \bar{\rho}_{a}\bar{\ga}_{a} + P_{a}v_{a}
\eea
We note that there is no central extension term. This is a departure from
\cite{Isidro1}.
$R_{gh}$ and $R_{agh}$ are the ghost no.operators. The total charge $R$ is
defined as
\be
R = R_{gh} - R_{agh}
\ee
The OPE of $R_{gh}$ and $R_{agh}$ with the other operators is as follows:
\bea
R_{gh}(z_{1})Q(z_{2}) &=& \frac{Q(z_{2})}{(z_{1}-z_{2})} \\
R_{agh}(z_{1}) Q(z_{2}) &=& 0  \\
R_{gh}(z_{1})\bar{Q}(z_{2}) &=& 0 \\
R_{agh}(z_{1})\bar{Q}(z_{2}) &=& \frac{\bar{Q}(z_{2})}{(z_{1}-z_{2})}  \\
R_{gh}(z_{1})G(z_{2}) &=& -\frac{G(z_{2})}{(z_{1}-z_{2})}  \\
R_{agh}(z_{1})\bar{G}(z_{2}) &=& -\frac{\bar{G}(z_{2})}{(z_{1}-z_{2})}  \\
T(z_{1})R_{gh}(z_{2}) &=& \frac{R_{gh}(z_{2})}{(z_{1}-z_{2})^2} +
\frac{\partial R_{gh}(z_{2})}{(z_{1}-z_{2})}  \\
T(z_{1})R_{agh}(z_{2}) &=& \frac{R_{agh}(z_{2})}{(z_{1}-z_{2})^2} +
\frac{\partial R_{agh}(z_{2})}{(z_{1}-z_{2})}
\eea
The OPE of $Q$ with $\bar{G}$ and $\bar{Q}$ with $G$ yields two operators $K$
and $\bar{K}$ which were studied by Hwang\cite{Hwang}. We find that
\bea
Q(z_{1})\bar{G}(z_{2}) &=& -\frac{\bar{K}(z_{2})}{(z_{1}-z_{2})^2}  \\
\bar{Q}(z_{1})G(z_{2}) &=& -\frac{K(z_{2})}{(z_{1}-z_{2})^2}
\eea
where
\bea
K &=& -{\rho}_{a}\bar{\ga}_{a} -
\frac{i}{2}f_{abc}v_{a}\bar{\ga}_{b}\bar{\ga}_{c} \\
\bar{K} &=& -\bar{\rho}_{a}\ga_{a} -\frac{i}{2}f_{abc}v_{a}\ga_{b}\ga_{c}
\eea
In \cite{Hwang}, Hwang showed that K is a solution to
\be
[K,[K,Q]] = 0
\ee
and is of the same form as (2.77).
It is easy to check that
\be
Q(z_{1})K(z_{2}) = \frac{\bar{Q}(z_{2})}{z_{1}-z_{2}}
\ee
while
\be
\bar{Q}(z_{1})\bar{K}(z_{2}) = \frac{Q(z_{2})}{z_{1}-z_{2}}
\ee
and
\bea
Q(z_{1})\bar{K}(z_{2}) &=& \bar{Q}(z_{1})K(z_{2}) = 0  \\
\eea
The operators $K$, $\bar{K}$, $G$ and $\bar{G}$ have the following OPE:
\bea
K(z_{1})G(z_{2}) &=& 0   \\
\bar{K}(z_{1})\bar{G}(z_{2}) &=& 0  \\
K(z_{1})\bar{G}(z_{2}) &=&  \frac{G(z_{2})}{z_{1}-z_{2}} +
\frac{V_{agh}}{(z_{1}-z_{2})^2}  \\
\bar{K}(z_{1})G(z_{2}) &=& \frac{\bar{G}(z_{2})}{z_{1}-z_{2}} -
\frac{V_{gh}(z_{2})}{(z_{1}-z_{2})^2}
\eea
where
\be
V_{gh} = v_{a}\ga_{a}
\ee
and
\be
V_{agh} = v_{a}\bar{\ga}_{a}.
\ee

{\em We thus have all the elements of an $N=4$ Superconformal Algebra}. However
due to the presence of the $V_{gh}$ and similar terms, it becomes infinite
dimensional as we shall see in the next section.
   We also have
\bea
R_{agh}(z_{1})G(z_{2}) &=& \frac{V_{agh}(z_{2})}{(z_{1}-z_{2})^2}  \\
R_{gh}(z_{1})\bar{G}(z_{2}) &=& -\frac{V_{gh}(z_{2})}{(z_{1}-z_{2})}
\eea
The only OPE where the central charge enters are as follows:
\bea
R_{gh}(z_{1})R_{gh}(z_{2}) &=& -\frac{d}{(z_{1}-z_{2})^2}  \\
R(z_{1})R(z_{2}) &=& -\frac{2d}{(z_{1}-z_{2})^2}
\eea
and
\bea
K(z_{1})\bar{K}(z_{2}) &=& \frac{d}{(z_{1}-z_{2})^2} -
\frac{R(z_{2})}{z_{1}-z_{2}}  \\
Q(z_{1})V_{agh}(z_{2}) &=& \frac{R_{agh}(z_{2})}{z_{1}-z_{2}} +
\frac{d}{(z_{1}-z_{2})^2}  \\
\bar{Q}(z_{1}V_{gh}(z_{2}) &=& -\frac{R_{gh}(z_{2})}{(z_{1}-z_{2})} +
\frac{d}{(z_{1}-z_{2})^2}
\eea
Also
\bea
 R_{gh}(z_{1})R_{gh}(z_{2}) &=& R_{agh}(z_{1})R_{agh}(z_{2}) =  0
\eea
The action of $K$ and $\bar{K}$ on $R_{gh}$ and $R_{agh}$ is as given by
\bea
R_{gh}(z_{1})K(z_{2}) &=& -\frac{K(z_{2})}{(z_{1}-z_{2})}  \\
R_{gh}(z_{1})\bar{K}(z_{2}) &=& \frac{\bar{K}(z_{2})}{z_{1}-z_{2}}  \\
R_{agh}(z_{1})K(z_{2}) &=& \frac{K(z_{2})}{z_{1}-z_{2}} \\
R_{agh}(z_{1}\bar{K}(z_{2}) &=& -\frac{\bar{K}(z_{2})}{z_{1}-z_{2}}
\eea
while
\bea
K(z_{1})K(z_{2}) &=& \bar{K}(z_{1}\bar{K}(z_{2}) = 0
\eea
We note that the operators $(K,\bar{K},-R)$ form an SU(2) algebra with $\bar{K}
\rightarrow T^{+}$, $K \rightarrow T^{-}$, and $R \rightarrow T^{3}$.
However, $K$ and $\bar{K}$ also behave as raising (lowering) operators wrt
$R_{agh}(R_{gh})$ respectively. This has interesting consequences in grouping
the various terms of the algebra into multiplets wrt $K$, $\bar{K}$ as we shall
see in Sec. 4. The OPE of $T$ with $K$ and $\bar{K}$ is as expected according
to the rules of conformal field theory:
\bea
T(z_{1})K(z_{2}) &=& \frac{K(z_{2})}{(z_{1}-z_{2})^2} +
\frac{\partial{K}(z_{2})}{z_{1}-z_{2}}  \\
T(z_{1})\bar{K}(z_{2}) &=& \frac{\bar{K}(z_{2})}{(z_{1}-z_{2})^2} +
\frac{\partial \bar{K}(z_{2})}{z_{1}-z_{2}}
\eea
\setcounter{equation}{0}
\sect{Proliferation of terms in the Algebra}
We now see how the algebra proliferates, actually becoming infinite
dimensional. Consider
\be
G(z_{1})V_{gh}(z_{1}) = \bar{G}(z_{1})V_{agh}(z_{2}) =
\frac{J_{V}(z_{2})}{z_{1}-z_{2}}
\ee
where
\be
J_{V}(z_{1}) = \frac{v_{a}J_{a}}{c_{A}}
\ee
Other operators of dimension 2 arise as follows. Consider
\be
G(z_{1})J_{V}(z_{1}) = \frac{P_{gh}(z_{2})}{(z_{1}-z_{2})^2} +
\frac{H(z_{2})}{z_{1}-z_{2}}
\ee
where
\bea
P_{gh} &=& \frac{v_{a}{\rho}_{a}}{c_{A}} \\
H &=& \frac{v_{a}{\rho}_{a}}{c_{A}} +
if_{abc}\frac{v_{a}{\rho}_{b}J_{c}}{c_{A}^2}
\eea
and
\be
\bar{G}(z_{1})J_{V}(z_{2}) = \frac{P_{agh}(z_{2})}{(z_{1}-z_{2})^2} +
\frac{H(z_{2})}{z_{1}-z_{2}}
\ee
where
\bea
P_{agh} &=& \frac{v_{a}{\rho}_{a}}{c_{A}} \\
\bar{H} &=& \frac{v_{a}\partial{R}_{a}}{c_{A}} +
if_{abc}\frac{v_{a}R_{b}J_{c}}{c_{A}^2}
\eea
Other operators of dimension two, not involving derivatives, arise in the OPE
of the following:
\bea
V_{gh}(z_{1})P_{gh}(z_{2}) &=& -\frac{{\cal V}(z_{2})}{z_{1}-z_{2}} \\
 V_{agh}(z_{1})P_{agh}(z_{2})&=& -\frac{{\cal V}(z_{2})}{z_{1}-z_{2}} \\
G(z_{1})\bar{G}(z_{2}) &=& \frac{W''(z_{2})-J_{V}(z_{2})}{(z_{1}-z_{2})^2} -
\frac{\partial{J_{V}}(z_{2})}{z_{1}-z_{2}} + \frac{W'(z_{2})}{z_{1}-z_{2}}
\eea
where
\bea
W''&=& \frac{R_{a}{\rho}_{a}}{c_{A}}   \\
{\cal V} &=& \frac{v_{a}v_{a}}{c_{A}}
\eea
and
\be
W' = \frac{R_{a}\partial {\rho}_{a}}{c_{A}} +
if_{abc}\frac{R_{a}{\rho}_{b}J_{c}}{c_{A}^2}
\ee
is a dimension 3 operator. We also note as in \cite{Isidro1} that
\be
G(z_{1})G(z_{2}) = \frac{W(z_{2})}{z_{1}-z_{2}}
\ee
where
\be
W = \frac{{\rho}_{a}\partial{\rho}_{a}}{c_{A}} +
if_{abc}\frac{{\rho}_{a}{\rho}_{b}J_{c}}{c_{A}^2}
\ee
and
\be
 \bar{G}(z_{1})\bar{G}(z_{2}) = \frac{\bar{W}(z_{2})}{z_{1}-z_{2}}
\ee
where
\be
\bar{W} = \frac{R_{a}\partial{R_{a}}}{c_{A}} +
if_{abc}\frac{R_{a}R_{b}J_{c}}{c_{A}^2}
\ee
Remarkably all the $W$s and $\bar{W}$s are closed wrt $Q$ and $\bar{Q}$,i.e.
\bea
Q(z_{1})W(z_{2}) &=& Q(z_{1})\bar{W}(z_{2}) = Q(z_{1})W'(z_{2}) = 0 \\
\bar{Q}(z_{1})W(z_{2}) &=& \bar{Q}(z_{2})\bar{W}(z_{2}) =
\bar{Q}(z_{1})W'(z_{2}) = 0
\eea
We also have the terms $V$, $\bar{V}$, which arise as follows:
\bea
Q(z_{1})V(z_{2}) &=& \frac{W(z_{1})}{z_{1}-z_{2}} \\
\bar{Q}(z_{1})\bar{V}(z_{2}) &=& \frac{\bar{W}(z_{2})}{z_{1}-z_{2}}
\eea
where
\bea
V = if_{abc}\frac{{\rho}_{a}{\rho}_{b}{\rho}_{c}}{c_{A}^2} \\
\bar{V} =
if_{abc}\frac{\bar{{\rho}}_{a}\bar{{\rho}}_{b}\bar{{\rho}}_{c}}{c_{A}^2}
\eea
According to expectation, we also see that
\bea
G(z_{1})W(z_{2}) &=& \frac{V(z_{2})}{(z_{1}-z_{2})^2}+
\frac{\partial{V}(z_{2})}{z_{1}-z_{2}}   \\
\bar{G}(z_{1})\bar{W}(z_{2}) &=& \frac{\bar{V}(z_{2})}{(z_{1}-z_{2})^2} +
\frac{\partial\bar{V}(z_{2})}{z_{1}-z_{2}}
\eea
We note also that $P_{agh}$ and $P_{gh}$ are closed wrt $Q(\bar{Q})$
respectively, while
\be
Q(z_{1})P_{gh}(z_{2}) = \frac{W''(z_{1})+J_{V}(z_{2})}{z_{1}-z_{2}} =
\bar{Q}(z_{1})P_{agh}(z_{2}) \
\ee
and
\bea
Q(z_{1})(W''(z_{2})+J_{V}(z_{2})) &=& \bar{Q}(z_{1})(W''(z_{2})+J_{V}(z_{2})) =
0  \\
G(z_{1})(W''(z_{2})+J_{V}(z_{2})) &=& \frac{2P_{gh}(z_{2})}{(z_{1}-z_{2})^2} +
\frac{\partial{P_{gh}(z_{2})}}{z_{1}-z_{2}}  \\
\bar{G}(z_{1})(W''(z_{2})+J_{V}(z_{2})) &=&
\frac{2P_{agh}(z_{2})}{(z_{1}-z_{2})^2} +
\frac{\partial{P_{gh}}(z_{2})}{z_{1}-z_{2}}
\eea
Other operators of dimension 3 arise as follows. Having noted that $K(\bar{K})$
are lowering(raising) operators wrt $R_{gh}$, we note that
\be
K(z_{1})\bar{V}_(z_{2}) = \frac{3V_{1/2}(z_{2})}{z_{1}-z_{2}} -
\frac{P_{agh}(z_{2})}{(z_{1}-z_{2})^2} +
\frac{\partial{P_{agh}(z_{2})}}{z_{1}-z_{2}} -
\frac{\bar{H}_{1}(z_{2})}{z_{1}-z_{2}}
\ee
where
\be
V_{1/2} = if_{abc}\frac{R_{a}R_{b}{\rho}_{c}}{3c_{A}^2}
\ee
and
\be
\bar{H_{1}} = \frac{v_{a}\partial{R_{a}}}{c_{A}}
\ee
Also
\be
\bar{K}(z_{1})V(z_{2}) = \frac{3V_{-1/2}(z_{2})}{z_{1}-z_{2}} -
\frac{P_{gh}(z_{2})}{(z_{1}-z_{2})^2} +
\frac{\partial{P_{gh}}(z_{2})}{z_{1}-z_{2}} - \frac{H_{1}(z_{2})}{z_{1}-z_{2}}
\ee
where
\be
V_{-1/2} = if_{abc}\frac{R_{a}{\rho}_{b}{\rho}_{c}}{3c_{A}^2}
\ee
and
\be
H_{1} = \frac{v_{a}\partial{{\rho}_{a}}}{c_{A}}
\ee
More dimension 3 operators are obtained if we consider the coefficients of the
quadratic terms in $(z_{1}-z_{2})^{-1}$ in the following:
\bea
J_{V}(z_{1})\bar{W}(z_{2}) &=& -\frac{X_{1}(z_{2})}{(z_{1}-z_{2})^2}  \\
J_{V}(z_{1})W'(z_{2}) &=& -\frac{X_{0}(z_{2})}{(z_{1}-z_{2})^2}  \\
J_{V}(z_{1})W(z_{2}) &=& -\frac{X_{-1}(z_{2})}{(z_{1}-z_{2})^2}
\eea
where
\bea
X_{1} &=& if_{abc}\frac{v_{a}R_{b}R_{c}}{c_{A}^2}  \\
X_{0} &=& if_{abc}\frac{v_{a}R_{b}{\rho}_{c}}{c_{A}^2}  \\
X_{-1} &=& if_{abc}\frac{v_{a}{\rho}_{b}{\rho}_{c}}{c_{A}^2}
\eea
Before we proceed further, we see first that the algebra does not close and we
end up with operators of higher and higher dimension. We take the example of
the following OPE:
\bea
J_{V}(z_{1})H(z_{2}) &=&
-if_{abc}\frac{\partial{v_{a}}v_{b}{\rho}_{c}(z_{2})}{c_{A}^2(z_{1}-z_{2})^2} +
if_{acd}if_{bec}\frac{v_{a}v_{b}{\rho}_{e}J_{d}(z_{2})}{c_{A}^2(z_{1}-z_{2})^2}
 \\
& = &\frac{X_{1/2}(z_{2})}{z_{1}-z_{2}}
\eea
We note that $X_{1/2}$ has dimension 4. If we take the OPE of $J_{V}$ with
$X_{1/2}$ again, we get an operator of dimension 5 due to the OPE of the $J$'s
with each other. This process of taking the OPE of the resulting operator with
$J_{V}$ can be repeated ad infinitum and each time we get an operator of with
the next higher dimension due to the fact that the OPE of a $J$ with a $J$
gives another $J$. Hence the algebra does not close and becomes infinite
dimensional.
\subsection{Form of Terms in Higher Dimensions}
What form do the operators in higher dimension take? To find out, we consider
the OPE of $X_{1/2}$ with $J_{V}$. We get
\be
J_{V}(z_{1})X_{1/2}(z_{2}) = \frac{X'_{-1/2}(z_{2})}{z_{1}-z_{2}} +
\frac{Y'_{-1/2}(z_{2})}{z_{1}-z_{2}}
\ee
where
\bea
X'_{-1/2} &=&
\frac{f_{abc}f_{dce}\partial{v_{e}}v_{d}v_{a}{\rho}_{b}}{c_{A}^3}\\
Y'_{-1/2} &=&
\frac{if_{abe}if_{cef}if_{dfg}v_{a}v_{c}v_{d}{\rho}_{b}J_{g}}{c_{A}^4}
\eea
  If the group is $SU(2)$, $if_{ijk} = i{\ep}_{ijk}$ and so, after a little
calculation, it turns out that
\be
X'_{-1/2} = \frac{1}{c_{A}}(P_{gh}\frac{\partial{\cal V}}{2}- {\cal
V}(\partial{P_{gh}}-H_{1}))
\ee
while,
\be
Y'_{-1/2} = \frac{1}{c_{A}}(\frac{{\cal V}}{c_{A}}(H-H_{1}))
\ee
We also note from (3.46) and (3.47) that as we go on to operators of higher and
higher dimension, we obtain operators which are tensors of the algebra of
successively higher rank. When the highest rank of the algebra has been
exhausted, the operators tend to take the form of products of these tensors. As
this part of the investigation is highly algebra dependent, we do not pursue it
further.
\subsection{Proliferation of Terms in Each Dimension}
So far, we have seen that the algebra is infinite dimensional if one considers
operators of all possible dimensions and is highly algebra dependent. However,
the no. of operators proliferates in each dimension.
   Let us group the operators that we have obtained so far in each dimension.
These are listed in Table 2.
\begin{table}
\caption{Operators in Each Dimension}
\begin{tabbing}
Dimension \=  Operator    \\
1 \> $Q, \bar{Q}, K, \bar{K}, R_{gh}, R_{agh}, V_{gh}, V_{agh}$   \\
2\>  $G, \bar{G}, T, J_{V}, P_{gh} P_{agh}, {\cal V}, W''$   \\
3\>  $W, W', \bar{W}, H, \bar{H}, X_{1}, X_{0}, X_{-1}, V, \bar{V}, V_{1/2},
V_{-1/2}, H_{1}, \bar{H_{1}} $
\end{tabbing}
\end{table}

\subsubsection{Action of $K(\bar{K})$ on ${\rho}$ and $R$}
We note the action of $K$ and $\bar{K}$ on ${\rho}_{a}$ and $R_{a}$. It is easy
to check that
\bea
K(z_{1}){\rho}_{a}(z_{2}) &=& 0 \\
\bar{K}(z_{1})R_{a}(z_{2}) &=& 0\\
K(z_{1})R_{a}(z_{2}) &=& \frac{{\rho}_{a}(z_{2})}{z_{1}-z_{2}} \\
\bar{K}(z_{1}){\rho}_{a}(z_{2})&=& \frac{R_{a}(z_{2})}{z_{1}-z_{2}}  \\
\eea
We also see that
\be
\bar{K}(z_{1})({\rho}_{d}(z_{2}){\rho}_{e}(z_{2})) =
\frac{R_{d}(z_{2}){\rho}_{e}(z_{2}) -
R_{e}(z_{2}){\rho}_{d}(z_{2})}{z_{1}-z_{2}}
\ee
and
\be
K(z_{1})(R_{d}(z_{2})R_{e}(z_{2})) =
\frac{{\rho}_{d}(z_{2})R_{e}(z_{2})-{\rho}_{e}R_{d}(z_{2})}{z_{1}-z_{2}}
\ee
while
\be
\bar{K}(z_{1}){\partial{\rho}_{d}(z_{2})} =
\frac{R_{d}(z_{2})}{(z_{1}-z_{2})^2} + \frac{\partial
R_{d}(z_{2})}{z_{1}-z_{2}}
\ee
and
\be
K(z_{1}){\partial{R}_{d}(z_{2})} = \frac{{\rho}_{d}(z_{2})}{(z_{1}-z_{2})^2} +
\frac{\partial{{\rho}}_{d}(z_{2})}{z_{1}-z_{2}}
\ee
Also
\be
K(z_{1})v_{a}(z_{2}) = \bar{K}(z_{1})v_{a}(z_{2}) = 0
\ee
Hence all terms of the form ${\rho}_{d}{\rho}_{e}$ must be accompanied by
$R_{d}{\rho}_{e}$ and $R_{d}R_{e}$, i.e. each ${\rho}$ is replaced by an $R$ ,
these being generated by the action of $K$ and $\bar{K}$ on the above.
    Similarly all terms of the form $\partial{\rho}_{d}$ must be accompanied by
$\partial{R}_{d}$ (within the same dimension) as well as an extra $R_{d}$ (in a
lower dimension).
  Finally, as
\bea
K(z_{1})\ga_{a}(z_{2}) &=& -\frac{\bar{\ga_{a}}(z_{2})}{z_{1}-z_{2}}\\
\bar{K}(z_{1})\bar{\ga}_{a}(z_{2}) &=& - \frac{\ga_{a}(z_{2})}{z_{1}-z_{2}}
\eea
while
\be
K(z_{1})\ga_{a}(z_{2}) = \bar{K}(z_{1})\ga_{a}(z_{2}) = 0
\ee
Hence all terms of the form $\ga_{a}\ga_{b}$ must be accompanied by
$\bar{\ga_{a}}\ga_{b}$ and $\bar{\ga_{a}}\bar{\ga}_{b}$ in order to close the
algebra.
 \subsubsection{Action of $V_{gh}(V_{agh})$ on ${\rho}(R)$:}
Next, we consider the action of $V_{gh}$ $(V_{agh})$ on a term like
${\rho}_{d}{\rho}_{e}(R_{d}R_{e})$. We get terms like
$(v_{d}{\rho}_{e}-v_{e}{\rho}_{d})(v_{d}R_{e}-v_{e}R_{d})$ which is effectively
replacing a ${\rho}(R)$ by its antisymmetric combination with $v$. Hence all
terms of the form ${\rho}_{d}{\rho}_{e}(R_{d}R_{e})$ are accompanied by terms
of the form $v_{d}{\rho}_{e}(v_{d}R_{e})$ in which a ${\rho}(R)$ is replaced by
a $v$.
\subsubsection{Action of $Q$}
$Q$ does not respect the ${\rho}{\leftrightarrow} R$ symmetry as
\be
Q(z_{1})v_{a}(z_{2}) = \frac{\bar{\rho}_{a}(z_{2})}{z_{1}-z_{2}}
\ee
and not $\bar{R_{a}}$.
Also
\be
Q(z_{1})\bar{\rho}_{a}(z_{2}) = 0
\ee
and
\be
Q(z_{1}){\rho}_{a}(z_{2}) = \frac{{\cal J}_{a}(z_{2})}{z_{1}-z_{2}}
\ee
as seen in Eq. (
and
\be
Q(z_{1})\partial{\rho}_{a}(z_{2}) =\frac{{\cal J}_{a}(z_{2})}{z_{1}-z_{2}} +
\frac{\partial{\cal J}_{a}(z_{2})}{z_{1}-z_{2}}
\ee
implying that {\em nonderivative terms in a higher dimension are given rise to
by derivative terms in a higher dimension}. Hence {\em all operators in a given
dimension cannot be obtained by the closure of the algebra in that dimension
alone}.
We now see how the no. of terms escalates. Consider, for example,
\be
Q(z_{1})H_{1}(z_{2}) = \frac{J_{V}(z_{2})}{(z_{1}-z_{2})^2} +
\frac{v_{d}\partial{J_{d}}}{c_{A}(z_{1}-z_{2})} +
\frac{{\rho}_{a}\partial{\rho}_{a}}{c_{A}(z_{1}-z_{2})} +
\frac{if_{abc}v_{a}\ga_{b}{\rho}_{c}}{(z_{1}-z_{2})^2} +
\frac{if_{abc}\partial({\rho}_{b}\ga_{c})v_{a}}{z_{1}-z_{2}}
\ee
Hence we end up with a term $f_{abc}v_{a}\ga_{b}{\rho}_{c}$ in dimension 2 in
addition to the new terms $v_{d}\partial{J_{d}}$,
$\bar{\rho}_{a}\partial{\rho}_{a}$ and
$if_{abc}\partial({\rho}_{b}\ga_{c})v_{b}$. The action of $Q$ on
$if_{abc}v_{a}\ga_{b}{\rho}_{c}$ yields an additional term $v_{a}\partial
\ga_{a}$ in dimension 2. However, the action of $Q$ on $v_{a}\partial\ga_{a}$
is as follows:
\be
Q(z_{1})(v_{a}\partial \ga_{a}(z_{2})) = \frac{R_{a}\partial
\ga_{a}}{z_{1}-z_{2}} + \frac{if_{abc}v_{a}\ga_{b}\ga_{c}}{(z_{1}-z_{2})^2}
\ee
so that we end up with an additional term in dimension 1, i.e.
$f_{abc}v_{a}\ga_{b}\ga_{c}$. We have not been able to find a general way of
inducing all terms,however, a list of some of the terms in dimension 1 is
presented in Table 3, accompanied by a similar list of dimension 2 terms in
Table 4. Both lists are, by no means exhaustive.
\begin{table}
\caption{Additional Operators in Dimension 1}
\begin{tabbing}
$if_{abc}v_{a}\ga_{b}\ga_{c}$\quad\quad\quad\quad \=
$if_{abc}\bar{\rho}_{a}\ga_{b}\ga_{c}$\quad\quad\quad\quad \=
$if_{abc}f_{def}f_{cdg}\ga_{a}\ga_{b}P_{e}v_{g}\bar\ga_{f}$ \\
$if_{abc}v_{a}\bar\ga_{b}\ga_{c}$\>$if_{abc}R_{a}\bar\ga_{b}\ga_{c}$\>
$if_{abc}f_{def}f_{cdg}\ga_{a}\ga_{b}P_{e}\bar{\rho}_{g}\bar\ga_{f}$ \\
$if_{abc}v_{a}\bar\ga_{b}\bar\ga_{c}$\>
$if_{abc}P_{a}v_{b}\ga_{c}$\>$if_{abc}f_{def}f_{ejk}f_{cdk}
\partial\bar\ga_{j}\ga_{a}\ga_{b}\bar\ga_{f}$ \\
$if_{abc}{\rho}_{a}\bar\ga_{b}\ga_{c}$\>
$if_{abc}\bar{\rho}_{a}\bar\ga_{b}\bar\ga_{c}$\\
$if_{abc}P_{a}\bar{\rho}_{b}\ga_{c}$\>$if_{abc}P_{a}v_{b}\ga_{c}$ \\
$if_{abc}R_{a}\bar\ga_{b}\bar\ga_{c}$\>$if_{abc}P_{a}R_{b}\bar\ga_{c}$
\end{tabbing}
\end{table}
\begin{table}
\caption{Additional Operators of Dimension 2}
\begin{tabbing}
$v_{a}\partial\ga_{a}$\quad\quad\quad \=
$v_{a}\partial\bar\ga_{a}$\quad\quad\quad \=
$if_{abc}\ga_{a}R_{b}{\rho}_{c}$\quad\quad\quad\=
$if_{abc}\bar\ga_{a}R_{b}R_{c}$ \\
$R_{d}\partial\ga_{d}$ \>
${\rho}_{d}\partial\ga_{d}$ \> $if_{abc}\bar\ga_{a}{\rho}_{b}{\rho}_{c}$ \>
$if_{abc}v_{a}\ga_{b}{\rho}_{c}$ \\
$R_{d}\bar\partial\ga_{d}$\>
${\rho}_{d}\partial\bar\ga_{d}$\>$if_{abc}v_{a}
\bar\ga_{b}{\rho}_{c}$\>$if_{abc}v_{a}\ga_{b}R_{c}$ \\
${\rho}_{d}\bar\partial\ga_{d}$\>$if_{abc}v_{a}
\bar\ga_{b}R_{c}$\>$if_{abc}v_{a}
\partial\bar\ga_{b}\ga_{c}$\>$if_{abc}v_{a}\partial\bar\ga_{b}\bar\ga_{c}$ \\
${\rho}_{a}\partial\ga_{a}+\partial{P_{a}}v_{a}$
\end{tabbing}
\end{table}
A complete analysis and classification of such terms is outside the scope of
the present paper.
\setcounter{equation}{0}
\sect{Analysis of our Results}
\begin{enumerate}
\item In our analysis of operators, we would like to introduce the terms,
operators of order one, two, three and so on. Operators of order one are
generated by the action of the operators $Q(\bar{Q})$ on operators of the same
dimension or lower. Operators of order two are generated by the action of
$Q(\bar{Q})$ on operators of one higher dimension. Operators of order three are
generated by the action of $Q(\bar{Q})$ on operators of dimension three.  The
list continues. We confine our discussion only to operators of order 1 with
dimensions upto three only.
It is interesting to group these operators into multiplets of $(K,\bar{K},
R_{gh})$. This is done in Table 5.
\begin{table}
\caption{Multiplets of $(K,\bar{K},R_{gh})$}
\begin{tabbing}
Dimension\quad\quad\quad\= Singlets\quad\quad\quad\=
Doublets\quad\quad\quad\=Triplets\quad\quad\quad\=Quadruplets \\
1\>    \> $Q, \bar{Q}$ \> $ K, R_{gh}, \bar{K}$ \>    \\
\>     \> $V_{gh}, V_{agh}$\> $\bar{K}, R_{agh}, K$\>   \\
2\> $W'', J_{V}$ \> $G$, $\bar{G}$\> \>  \\
\>  $ {\cal V}$ \> $P_{gh}$, $P_{agh}$\>  \>   \\
3\>   \> $H$,$\bar{H}$ \>$W$,$W'$,$W''$\>$\bar{V}$, $V_{-1/2}$,$V_{1/2}$,$V$ \\
\>  \>  $H_{1}$,$\bar{H}_{1}$ \> \>
\end{tabbing}
\end{table}
The BRST structure of some of these operators is illustrated in Table 6. Their
anti BRST structure is illustrated in Table 7.
\begin{table}
\caption{BRST structure of the Operators}
\begin{tabular}{clllll}
Dimension&Doublets\\
1& $(Q, R_{gh})$& $(R_{agh}, V_{agh})$& $(\bar{Q}, K)$& $(\bar{K}, V_{gh})$&\\
2& $(W''+J_{V}, P_{gh})$& \hspace{-2mm}$(P_{agh}, {\cal V})$&
\hspace{-8mm}$(-\bar{G}+V_{gh}, J_{V})$ &$(\bar{G}-V_{gh},W'')$&$(T+R_{gh},G) $
\\
3& $(W,V)$& $(W',H)$& $(\bar{W},\bar{H})$&$(\bar{V},X_{1})$ &
\end{tabular}
\end{table}

\begin{table}
\caption{Anti BRST structure of the Operators}
\begin{tabular}{clllll}
Dimension&Doublets\\
1& $(\bar{Q}, R_{agh})$& $(R_{gh}, V_{gh})$& $(Q, \bar{K})$& $(K, V_{agh}) $ \\
2& $(W''+J_{V}, P_{agh})$& $(P_{gh}, {\cal V})$& \hspace{-31mm}$(-G+V_{agh},
J_{V})$ &\hspace{-20mm}$(G-V_{gh},W'')$&\hspace{-12mm}$(T+R_{agh},\bar{G}) $ \\
3& $ (\bar{W},\bar{V})$& $(-W'+\partial{W''}-\partial{J_{V}},\bar{H})$&
$(-W,H)$&$(V,X_{-1})$
\end{tabular}
\end{table}
      The doublet structure of the BRST and anti BRST multiplets (i.e. absence
of BRST and anti BRST invariant, non-exact operators) is highlighted as is the
duality between the BRST and anti BRST structures, in spite of the vastly
differing form of the BRST and anti BRST operators.
\item We now look at the algebra of the operators with $v=P=0$. We end up with
the operators given in Table 8.
\begin{table}
\caption{Operators with $v=P=0$}
\begin{tabular}{cl}
Dimension &Operators\\
1& $Q=-\ga_{a}J_{a} -\frac{if_{abc}}{2}\ga_{a}\ga_{b}{\rho}_{c}$\\
 & $\bar{Q} = -\bar\ga_{a}J_{a}-
\frac{if_{abc}}{2}\bar\ga_{a}\bar\ga_{b}\bar{\rho}_{c}-if_{abc}
\bar\ga_{a}\ga_{b}{\rho}_{c}$ \\
 & $R_{gh}={\rho}_{a}\ga_{a}$\\
 &$R_{agh}= \bar{\rho}_{a}\bar\ga_{a}$  \\
2& $T=
{\rho}_{a}\partial\ga_{a}+\bar{\rho}_{a}\partial\bar\ga_{a}-
\frac{J_{a}J_{a}}{c_{A}}$ \\
 & $G=-\frac{{\rho}_{a}J_{a}}{c_{A}}$\\
 &  $\bar{G}=-\frac{\bar{\rho}_{a}J_{a}}{c_{A}}$\\
 & $W''= \bar{\rho}_{a}{\rho}_{a}$    \\
3&  $ W=\frac{{\rho}_{a}\partial{\rho}_{a}}{c_{A}} +
if_{abc}\frac{{\rho}_{a}{\rho}_{b}J_{c}}{c_{A}^2}$ \\
 & $W'= \frac{\bar{\rho}_{a}\partial{\rho}_{a}}{c_{A}} +
if_{abc}\frac{{\rho}_{a}\bar{\rho}_{b}J_{c}}{c_{A}^2}$ \\
 & $\bar{W}= \frac{\bar{\rho}_{a}\partial\bar{\rho}_{a}}{c_{A}} +
if_{abc}\frac{\bar{\rho}_{a}\bar{\rho}_{b}J_{c}}{c_{A}^2}$ \\
 & $V=if_{abc}\frac{{\rho}_{a}{\rho}_{b}{\rho}_{c}}{3c_{A}^2}$ \\
 & $V_{-1/2}=if_{abc}\frac{{\rho}_{a}{\rho}_{b}\bar{\rho}_{c}}{3c_{A}^2}$ \\
 & $V_{1/2}=if_{abc}\frac{{\rho}_{a}\bar{\rho}_{b}\bar{\rho}_{c}}{3c_{A}^2}$ \\
  &
$\bar{V}=if_{abc}\frac{\bar{\rho}_{a}\bar{\rho}_{b}\bar{\rho}_{c}}{3c_{A}^2}$
\\
4& $X_{-
1/2}=if_{adc}\frac{\partial{\rho}_{a}{\rho}_{d}
\bar{\rho}_{e}}{c_{A}^2}+f_{def}f_{afg}
\frac{{\rho}_{a}{\rho}_{d}\bar{\rho}_{e}J_{g}}{c_{A}^3}$ \\
 &
$X_{1/2}=if_{adc}\frac{\partial{\rho}_{a}
\bar{\rho}_{d}\bar{\rho}_{e}}{c_{A}^2}+
f_{def}f_{afg}\frac{{\rho}_{a}\bar{\rho}_{d}\bar{\rho}_{e}J_{g}}{c_{A}^3}$\ \
\end{tabular}
\end{table}
In this case, $R_{a}=\bar{\rho}_{a}$ and the algebra is seen to truncate at
dimension 4. $K$ and $\bar{K}$ do not act any more as SU(2) raising and
lowering operators and all relevence to antighosts drop out of the BRST
operator when $P_{a}$ is put to zero.
 The central term reappears in the OPE of $T$ with $T$ as follows:
\be
T(z_{1})T(z_{2}) = \frac{-d}{(z_{1}-z_{2})^4} +
\frac{{2T}(z_{2})}{(z_{1}-z_{2})^2}+\frac{\partial{T}(z_{2})}{z_{1}-z_{2}}
\ee
The OPE of $Q$ with $G$ does not yield the total energy momentum tensor and the
central term reappears as follows:
\be
Q(z_{1})G(z_{2}) = \frac{d}{(z_{1}-z_{2})^3} -
\frac{J_{a}J_{a}(z_{2})}{c_{A}(z_{1}-z_{2})} +
\frac{{\rho}_{a}\partial\ga_{a}(z_{2})}{z_{1}-z_{2}}+
\frac{{\rho}_{a}\ga_{a}(z_{2})}{(z_{1}-z_{2})^2}
\ee
Clearly if antighosts are to be introduced, the Lagrange multipliers cannot be
discarded as they enter into the BRST operator via the term $P_{a}{\rho}_{a}$.
If all antighosts are put equal to zero along with the $v$'s and the $P$'s, we
recover the algebra of \cite{Isidro1}.
\item It is instructive to compare the above algebra with that of
\cite{Isidro1} and \cite{Figueroa}. In Ref. \cite{Figueroa}, the OPE of $G$
with $G$ is zero, while in \cite{Isidro1} it is not, being given by Eq.(3.15).
however, if one considers the complete expression for the operator $G$ in
\cite{Isidro1}, as given in Eq. (2.50), we note that it forms a representation
for the SU(2) triplet of operators $(K,\bar{K},R_{gh})$. This is not so for the
operator $G$ in Ref. \cite{Figueroa}, because the term
$\frac{1}{2}f_{ijk}{\rho}_{i}{\rho}_{j}\ga_{k}$ transforms differently from the
others under the action of $K(\bar{K})$. We were unable to find an expression
for $G(\bar{G})$: $(G \bar{G})$ was a multiplet of $(K \bar{K} R_{gh})$ and
$GG=\bar{G}\bar{G} = 0 $. In that case, the algebra would truncate and many
terms would drop out.
\item We now take a look at the BRST and anti BRST currents and how they
transform under the action of the principal operators in the algebra i.e. \\
$(Q,G,\bar{Q},\bar{G},T,K,\bar{K},R_{gh},R_{agh})$. After a little calculation
it can be shown that
\bea
Q(z_{1}){\rho}_{a}(z_{2}) &=& \frac{{\cal J}_{a}(z_{2})}{z_{1}-z_{2}} \\
Q(z_{1}){\cal J}_{a}(z_{2}) &=& 0    \\
G(z_{1}){\rho}_{a}(z_{2}) &=&  0   \\
G(z_{1}){\cal J}_{a}(z_{2}) &=& \frac{{\rho}_{a}(z_{2})}{(z_{1}-z_{2})^2} +
\frac{\partial{\rho}_{a}(z_{2})}{z_{1}-z_{2}} \\
\bar{Q}(z_{1})\bar{\rho}_{a}(z_{2}) &=& \frac{{\cal T}_{a}(z_{2})}{z_{1}-z_{2}}
\\
\bar{Q}(z_{1}){\cal T}_{a}(z_{2}) &=& 0    \\
\bar{G}(z_{1})\bar{\rho}_{a}(z_{2}) &=&  0   \\
\bar{G}(z_{1}){\cal T}_{a}(z_{2}) &=&
\frac{\bar{\rho}_{a}(z_{2})}{(z_{1}-z_{2})^2} +
\frac{\partial\bar{\rho}_{a}(z_{2})}{z_{1}-z_{2}} \\
Q(z_{1})\bar{\rho}_{a}(z_{2}) &=&0  \\
Q(z_{1}){\cal T}_{a}(z_{2}) &=& 0    \\
G(z_{1})\bar{\rho}_{a}(z_{2}) &=& \frac{v_{a}(z_{2})}{(z_{1}-z_{2})^2} +
\frac{\partial{v}_{a}(z_{2})}{z_{1}-z_{2}} \\
G(z_{1}){\cal T}_{a}(z_{2}) &=& \frac{\bar R_{a}(z_{2})}{(z_{1}-z_{2})^2} +
\frac{\partial{\bar R}_{a}(z_{2})}{z_{1}-z_{2}} \\
\bar{Q}(z_{1}){\rho}_{d}(z_{2}) &=&
\frac{if_{dac}\bar\ga_{a}(z_{2}){\rho}_{c}(z_{2})}{z_{1}-z_{2}}  \\
\bar{Q}(z_{1}){\cal J}_{d}(z_{2}) &=& \frac{if_{dac}\bar\ga_{a}(z_{2}){\cal
J}_{c}(z_{2})}{z_{1}-z_{2}}+
\frac{if_{dae}{\rho}_{a}(z_{2})P_{e}(z_{2})}{z_{1}-z_{2}}  \\
\bar{G}(z_{1}){\rho}_{d}(z_{2}) &=&
\frac{if_{dac}J_{a}(z_{2})v_{c}(z_{2})}{z_{1}-z_{2}}-
\frac{v_{d}(z_{2})}{(z_{1}-z_{2})^2}-
\frac{\partial{v}_{d}(z_{2})}{z_{1}-z_{2}} \\
\bar{G}(z_{1}){\cal J}_{d}(z_{2}) &=&
\frac{\bar{\rho}_{d}(z_{2})}{(z_{1}-z_{2})^2} +
\frac{\partial{R}_{d}(z_{2})}{z_{1}-z_{2}} +
\frac{if_{dac}R_{a}(z_{2})J_{c}(z_{2})}{z_{1}-z_{2}}-  \non \\ &&
\frac{if_{dac}\partial{v}_{a}(z_{2})\ga_{c}(z_{2})}{z_{1}-z_{2}}
\eea
where
\be
\bar{R}_{a} = {\rho}_{a} + if_{abc}v_{b}\bar\ga_{c}
\ee
 We note that the topological BRST charges $(Q,G)$ and the topological currents
$({\rho}_{a},{\cal J}_{a})$ are compatible, as are the topological anti BRST
charges $(\bar{Q},\bar{G})$ and the topological anti BRST currents $
(\bar{\rho}_{a}, {\cal T}_{a})$. The topological anti BRST currents are
invariant under the action of the topological BRST charges. However, this does
not hold for the BRST currents under the action of the topological anti BRST
charges as is also demonstrated by the action of $\bar{G}$ on ${\cal J}_{a}$
which has terms in both $\bar{\rho}_{a}$ and $R_{a}$, in contrast with
eq.(4.14) where only $\bar{R_{a}}$ enters. The reason is probably due to the
fact that the anti BRST current is more complete with more terms in order to be
compatible with the BRST transformation.
\item We close this section with a comment on the existence of two independent
currents that close an affine algebra with levels $k_{1}$ and $k_{2}$
respectively as in Ref.\cite{Isidro1}. The total current ${\cal
J}_{a}=J^{1}_{a}+J^{2}_{a}+if_{abc}\ga_{b}{\rho}_{c}$ and if we assume that
$J^{1}_{a}$ and $J^{2}_{a}$ transform under the BRST symmetry as in Eq.(2.13) ,
we end up as in Ref.\cite{Isidro1} with the following condition for the sum of
the levels $k_{1}$ and $k_{2}$:
\be
k_{1}+k_{2} = c_{A}
\ee
As in \cite{Isidro1}, only the sum of the current algebra levels is
constrained. Similarly, using Eq.(2.29 ), we see that the anti BRST variation
of the currents restricts the sum of the levels $k_{1}+k_{2}$ to
\be
k_{1}+k_{2} = - c_{A}
\ee
in order to achieve nilpotency for the anti BRST transformation.
As in Ref. \cite{Isidro1}, denoting $k_{1}$ by $k$, we get
\be
T= \frac{1}{2k+c_{A}}(J_{a}^{1}J_{a}^{1} - J_{a}^{2}J_{a}^{2})
+{\rho}_{a}\partial\ga_{a} + \bar{\rho}_{a}\partial \bar\ga_{a} +
v_{a}\partial{P_{a}}
\ee
where we applied the BRST transformation on
\be
G = \frac{1}{2k+c_{A}}:[{\rho}_{a}(J_{a}^{1}-J_{a}^{2}): +
if_{abc}:v_{b}\ga_{c}(J_{a}^{1}-J_{a}^{2}) -v_{a}\partial\ga_{a}]
\ee
thus seeing  that $T$ is both BRST and anti BRST exact. The currents
$J_{a}^{1}$ and $J_{a}^{2}$ contribute $\frac{2kdimg}{2k+c_{A}}$ and
$-\frac{2k+2c_{A}}{2k+c_{A}}dimg$ respectively whose sum exactly cancels the
central charge of the ghost system. We see that the system offers a description
of the gauged WZNW model.
As noted in \cite{Isidro1}, the construction cannot be extended to three or
more currents.
\end{enumerate}
\setcounter{equation}{0}
\sect{Discussion and Conclusion}
 We thus see that the full topological algebra involving ghosts, antighosts and
free multipliers becomes indefinitely large for each operator dimension. This
is possibly the price one pays for the disappearance of the central term in the
OPE of $Q$ with $G$ and $\bar{Q}$ with $\bar{G}$. We still have an $N=4$
Superconformal Algebra, albeit infinite dimensional.
 The ghost no.(antighost no.) charges  $R_{gh}(R_{agh})$ have been redefined
from ${\rho}_{a}\ga_{a}$ to ${\rho}_{a}\ga_{a} + P_{a}v_{a}$ and
$\bar{\rho}_{a}\bar\ga_{a}$ to $\bar{\rho}_{a}\bar\ga_{a} + P_{a}v_{a}$. The
$P_{a}v_{a}$ term acts as a sort of gauge transformation term. The central term
of the affine algebra arises only in the OPE of the $SU(2)$ charges
$(K,\bar{K},R_{gh},R_{agh})$ with their opposites.
   The doublet structure of the operators under the action of the BRST and the
anti BRST charges is also noted.
   It is also pointed out that the algebra is finite when the parameters
$v_{a}$ and $P_{a}$ are put to zero. However, the central term reappears in the
OPE of $T$ with $T$ and that of $Q$ with $G$ and $\bar{Q}$ with $\bar{G}$.
Moreover, the energy momentum tensor appearing in the OPE of  $Q$ with $G$ and
$\bar{Q}$ with $\bar{G}$ is not complete, since the antighost(ghost)
contribution is missing.
   As seen in the last section, the algebra of the topological BRST and the
anti BRST charges and the topological BRST and the anti BRST currents
$(Q,G,{\rho}_{a},{\cal J}_{a})$ and $(\bar{Q},\bar{G},\bar{\rho}_{a},{\cal
T}_{a})$ are compatible. The anti BRST current is also invariant under the
action of the BRST charge, however, the same does not hold for the action of
the anti BRST charge on the BRST current. The two sets of algebras are thus not
compatible with each other.
   Finally, we note that for two independent affine currents $J_{1}^{a}$ and
$J_{2}^{a}$ with levels $k_{1}$ and $k_{2}$ : $k_{1}+k_{2}=c_{A}$, the above
construction holds true, as the BRST charge contains the sum $(J_{1}^{a}
+J_{2}^{a})$. Results for the topological algebra obtained above can thus be
applied to the gauged WZNW model.

\subsection*{Acknowledgements}
It is a pleasure to thank Stephen Hwang and Henric Rhedin for numerous
discussions on all aspects of the manuscript and especially help in calculating
complicated OPE's. I would also like to thank Martin Cederwall for discussions
on the structure of the $N=4$ SCA and Niclas Wyllard, Stefan Rommer and Anders
Westerberg for help with the \LaTeX typing of the manuscript. E-mail
correspondence with Kris Thielemans, J.Isidro and A.Ramallo is also gratefully
acknowledged.


\begin{thebibliography}{10}

\bibitem{Witten1}
E. Witten, Comm. in Math. Phys. {\bf 117} (1988) 353;

\bibitem{Birmingham}
D. Birmingham, M. Blau, M. Rakowski, G.Thompson, Phys. Rep. {\bf 209}(1991)
129;

\bibitem{Witten2}
E. Witten, Nucl. Phys. {\bf B340} (1990) 281;

\bibitem{Dijkgraaf}
R. Dijkgraaf, E.Verlinde and H.Verlinde, Nucl. Phys. {\bf B352} (1991) 59;

\bibitem{Spiegelglas}
M. Spiegelglas and S.Yankielowicz, Nucl. Phys.  {\bf B393} (1993) 301; O.
Aharony et al, Nucl. Phys. {\bf B399} (1993) 527; {\em ibid}, Phys. Lett. {\bf
B289} (1992) 309; {\em ibid}, {\bf B305} (1993) 35;

\bibitem{Hu}
H. L. Hu and M. Yu, Phys. Lett. {\bf B289} (1992) 302; {\em ibid}, Nucl. Phys.
{\bf B391} (1993) 389;

\bibitem{Sadov}
V. Sadov, Int. Journal of Mod. Phys. {\bf A8} (1993) 5115;

\bibitem{Isidro1}
 J. Isidro and A. V. Ramallo , Phys. Lett. {\bf B316} (1993)  488;

\bibitem{Figueroa}
J. M. Figueroa O'Farrill, Phys. Lett. {\bf B316} (1993) 496;

\bibitem{Isidro2}
J. Isidro and A.V.Ramallo, Phys. Lett. {\bf B340} (1994) 48;

\bibitem{Ennes}
I. P.Ennes, J. Isidro and A. V. Ramallo, {\em The BRST Symmetry of Affine Lie
Superalgebras and non-critical strings } , hep-th/ 9508030;

\bibitem{Kazama}
Y. Kazama, Mod. Phys. Lett. {\bf A6} (1991) 1321;

\bibitem{Eguchi}
T. Eguchi and S. K. Yang, Mod. Phys. Lett. {\bf A4} (1990) 1653; T.Eguchi, S.
K. Hosono and S.K.Yang, Comm. Math. Phys., {\bf 140} (1991) 159;

\bibitem{Curci}
G. Curci and R. Ferrari, Phy. Lett. {\bf B63} (1976) 91;

\bibitem{Ojima}
I.Ojima, Progr. Theor. Phys., {\bf 64} (1980) 625;

\bibitem{Baulieu}
L. Baulieu and J. Thierry-Mieg, Nucl. Phys. {\bf B197} (1982) 477;

\bibitem{Hwang}
S. Hwang, Nucl. Phys., {\bf B231} (1984) 386; {\em ibid} , Nucl. Phys. {\bf
B322} (1989) 107;

\bibitem{Perry}
M. J. Perry and E. Teo, Nucl. Phys., {\bf B392} (1993) 369;

\bibitem{Preitschopf}
C. Preitschopf, M. Weinstein, D. Nemeschansky, Ann. of Phys.,{\bf
183}(1988)226;

\bibitem{Kulshreshtha}
Usha Kulshreshtha, D. S. Kulshreshtha and H. J. Muller Kristen, Can. Journal of
Phys.,{\bf 72}(1994) 639(and references therein);

\bibitem{Fulop}
G. F\"ul\"op, Ph. D Thesis (G\"{o}teborg University) (1996)

\bibitem{Witten3}
E. Witten, Comm. of Math. Phys. {\bf 92} (1984)455;

\bibitem{Fuchs}
J.Fuchs, {\em Affine Lie Algebras and Quantum Groups} ,Oxford University Press
(1994 )\
\bibitem{Kar} \
D. S. Karabali and H. J.Schnitzer, Nucl. Phys. {\bf B329}(1989)649; S. Hwang
and H. Rhedin, Nucl. Phys. {\bf B406} (1993)165;

\end{thebibliography}
\end{document}